\documentclass[10pt,conference]{IEEEtran}
\IEEEoverridecommandlockouts
\usepackage[T2A]{fontenc}
\usepackage{xcolor}
\usepackage{arydshln}
\usepackage{hyperref}
\usepackage{enumitem}
\usepackage{tcolorbox}
\usepackage{caption}
\usepackage{subcaption}
\usepackage{placeins}
\usepackage{arydshln}
\usepackage{CJKutf8}
\usepackage{multirow}
\usepackage{pythonhighlight}
\usepackage[T2A,T1]{fontenc}
\usepackage[normalem]{ulem}
\usepackage{pifont}
\newcommand{\cmark}{\ding{51}}%
\newcommand{\xmark}{\ding{55}}%
\newcommand{\finding}[2]{
\begin{tcolorbox}[width=\linewidth,colback=black!5!white,colframe=black!75!black]
\textbf{Findings #1:} 
{#2}
\end{tcolorbox}}

\definecolor{mygray}{gray}{0.8}
\usepackage{cite}
\usepackage{amsmath,amssymb,amsfonts}
\usepackage{algorithmic}
\usepackage{graphicx}
\usepackage{textcomp}
\usepackage{xcolor}
\usepackage{filecontents}
\def\BibTeX{{\rm B\kern-.05em{\sc i\kern-.025em b}\kern-.08em
    T\kern-.1667em\lower.7ex\hbox{E}\kern-.125emX}}
\makeatletter 
\newcommand{\linebreakand}{%
  \end{@IEEEauthorhalign}
  \hfill\mbox{}\par
  \mbox{}\hfill\begin{@IEEEauthorhalign}
}
\makeatother 
\begin{document}

\title{Pop Quiz! Do Pre-trained Code Models Possess
Knowledge of Correct API Names?
}
\author{\IEEEauthorblockN{Terry Yue Zhuo}
\IEEEauthorblockA{\textit{Monash University,} \\
\textit{CSIRO's Data61}\\
\textit{Australia}\\
terry.zhuo@monash.edu}
\and
\IEEEauthorblockN{Xiaoning Du}
\IEEEauthorblockA{\textit{Monash University} \\
\textit{Australia}\\
xiaoning.du@monash.edu}
\and
\IEEEauthorblockN{Zhenchang Xing}
\IEEEauthorblockA{\textit{CSIRO's Data61,} \\
\textit{Australian National University} \\
\textit{Australia}\\
zhenchang.xing@data61.csiro.au}
\and
\IEEEauthorblockN{Jiamou Sun}
\IEEEauthorblockA{\textit{CSIRO's Data61} \\
\textit{Australia}\\
frank.sun@data61.csiro.au}
\linebreakand
\IEEEauthorblockN{Haowei Quan}
\IEEEauthorblockA{\textit{Monash University} \\
\textit{Australia}\\
haowei.quan@monash.edu}
\and
\IEEEauthorblockN{Li Li}
\IEEEauthorblockA{\textit{Beihang University} \\
\textit{China}\\
lilicoding@ieee.org}
\and
\IEEEauthorblockN{Liming Zhu}
\IEEEauthorblockA{\textit{CSIRO's Data61,} \\
\textit{University of New South Wales}\\
\textit{Australia}\\
liming.zhu@data61.csiro.au}
}
\maketitle
\begin{abstract}
Recent breakthroughs in pre-trained code models, such as CodeBERT and Codex, have shown their superior performance in various downstream tasks. 
The correctness and unambiguity of API usage among these code models are crucial for achieving desirable program functionalities, requiring them to learn various API fully qualified names structurally and semantically. 
Recent studies reveal that even state-of-the-art pre-trained code models struggle with suggesting the correct APIs during code generation.
However, the reasons for such poor API usage performance are barely investigated.
To address this challenge, we propose using knowledge probing as a means of interpreting code models, which uses cloze-style tests to measure the knowledge stored in models. Our comprehensive study examines a code model's capability of understanding API fully qualified names from two different perspectives: API call and API import.  Specifically, we reveal that current code models struggle with understanding API names, with pre-training strategies significantly affecting the quality of API name learning. We demonstrate that natural language context can assist code models in locating Python API names and generalize Python API name knowledge to unseen data. Our findings provide insights into the limitations and capabilities of current pre-trained code models, and suggest that incorporating API structure into the pre-training process can improve automated API usage and code representations. This work provides significance for advancing code intelligence practices and direction for future studies. All experiment results, data and source code used in this work are available at \url{https://doi.org/10.5281/zenodo.7902072}.

\end{abstract}
\section{Introduction}

Recent advances in code intelligence have incorporated pre-training techniques, where models are pre-trained on large-scale unlabeled source code corpora to learn the code's representation and semantics. The pre-trained code models, such as CodeBERT~\cite{feng2020codebert} and GraphCodeBERT~\cite{guo2021graphcodebert}, can be fine-tuned for various downstream code tasks, such as code completion and translation~\cite{allamanis2018survey,10.1145/3487569,niudeep}.

Despite the improvement, there is still a significant performance gap between their performance and that of human developers when it comes to using APIs correctly. For instance, several studies have demonstrated that even state-of-the-art pre-trained code models, such as Codex~\cite{chen2021evaluating}, StarCoder~\cite{li2023starcoder} and GPT-4~\cite{openai2023gpt4}, struggle with suggesting the correct APIs during code generation~\cite{lai2022ds,yu2023codereval}. However, few studies have investigated the reasons behind these models' poor API usage performance.

Basically, correct API usage depends on two types of knowledge: how to invoke an API and which API to invoke, with the former being a fundamental step towards the latter. To invoke an API, one must have knowledge of the code grammar of importing libraries and composing the correct API name based on the import and call statements. However, code models are not explicitly instructed by code grammar. Although we often assume that the models can learn to use APIs effectively by observing a large number of examples, this assumption has been poorly validated. Therefore, we pose the following question: \textbf{Do Pre-trained Code Models Possess Knowledge of Correct API Names?}

The fully qualified name of an API comprises not only the function name but also the name of the package, module, and/or class to which it belongs.
To ease the presentation, we will use the term ``module'' to refer to these entities below.
Libraries usually organize APIs into nested modules to help developers understand the available features when using the APIs.
APIs are typically organized into nested modules within libraries to aid developers in understanding the available features when using the APIs. These modules form a hierarchical structure, with the higher-level module serving as the parent of its direct lower-level modules, and the APIs acting as the leaf nodes in the structure. The fully qualified API names are obtained by traversing the hierarchy from the root to the leaf, with all names connected by dots.
Depending on the imported module's level and whether it is associated with an alias, the API name used for invocation should be adjusted accordingly. This convention can be easily understood by human developers, but code models may find it challenging to learn.

\begin{figure}
    \centering
    \includegraphics[width=\columnwidth]{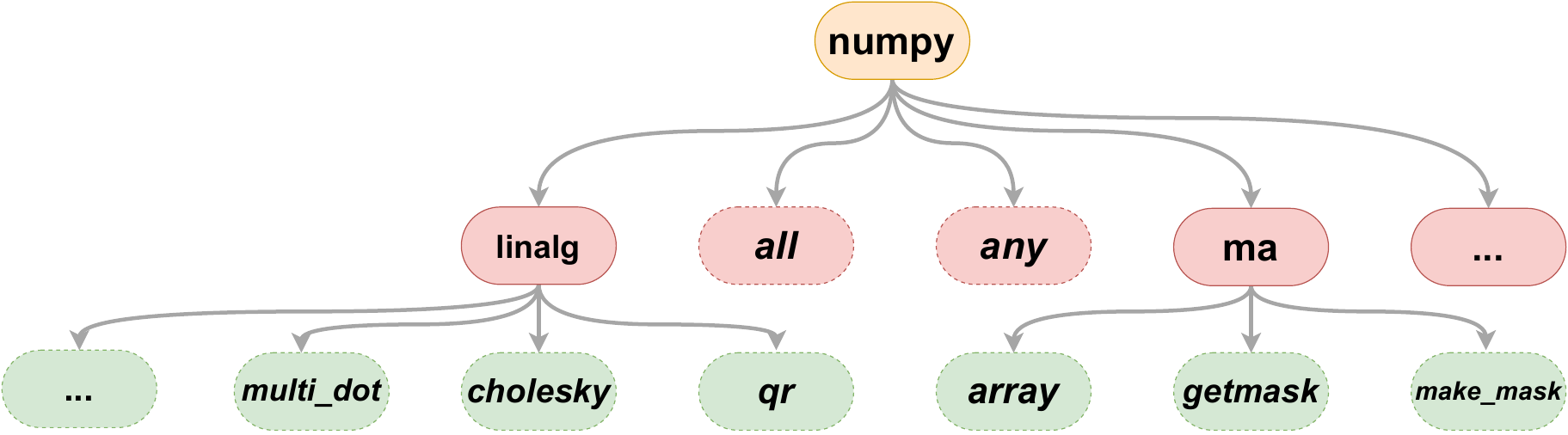}
    \caption{An example of module design of \texttt{<numpy>} library. 
    }
    \label{fig:intro}
\end{figure}

In addition, the fully qualified names of APIs convey information about code modularization and API namespace design. If models can learn a good representation of API names, they could be useful for automating API namespace design by providing more precise and relevant design options.
For example, consider the Python library \texttt{<numpy>}, which is used for mathematical operations on arrays and matrices. 
As shown in Figure~\ref{fig:intro}, \texttt{<numpy>} includes several modules, including \texttt{<linalg>} (short for linear algebra) for APIs related to computations in linear algebra, such as \texttt{<multi\_dot>} (dot product), \texttt{<cholesky>} (Cholesky decomposition), and \texttt{<qr>} (QR factorization), as well as \texttt{<ma>} (short for masked array) for APIs related to operations on masked arrays. 
Such namespace design can provide valuable insights when designing relevant libraries or similar libraries for other programming languages.

To understand how well pre-trained code models understand API names, it is essential to analyze and interpret the internal mechanisms of pre-trained code models. However, deep neural models are generally complex and opaque~\cite{samek2017explainable}, making it hard to fully understand their internal mechanisms. 
In the natural language processing community, the mainstream technique to examine what pre-trained language models know is probing, where a probe is a simple classifier that takes the model’s contextual representations as input and predicts a property of interest~\cite{jiang2020can}.
Recently, researchers have attempted to analyze the code syntax captured by code models via attention analysis and grammatical structure probing~\cite{wan2022they,hernandez2022ast,troshin2022probing}. 
However, there is an insubstantial effort on assessing API knowledge~\cite{huang2022prompt}.

In this work, we explore the interpretability of code models and introduce a large-scale knowledge probing task that specifically targets fully qualified names of APIs. 
We design an automated and effective evaluation framework, \texttt{INK}, to probe code models' understanding of API names with cloze-style pop quizzes.
To generate the quizzes, we first extract API full qualified names from API calls in large-scale source code corpora that are commonly used for training code models.
Note that, instead of directly extracting API names from libraries, we take into account the limitations of code models during their training. 
We refrain from testing the models on the knowledge they have never been exposed to.
Further, we derive import statements and API call statements based on the API names and mask out different tokens, with only one token masked out at a time, at each module level of these statements. For instance, when invoking \texttt{<multi\_dot>} module as shown in Figure~\ref{fig:intro}, we consider from two aspects, API call (\texttt{<numpy.linalg.multi\_dot>}) and API import (\texttt{<from numpy.linalg import multi\_dot>}). By masking token \texttt{lin}  by ``\texttt{[MASK]}'', we get two quizzes of \texttt{<numpy.[MASK]alg.multi\_dot>} and \texttt{<from numpy.[MASK]alg import multi\_dot>}.
The code models are then expected to predict the masked token given an import or API call statement with the mask applied. 
To better understand the capabilities and limitations of pre-trained code models in learning API names, we set out to investigate several research questions (RQs) to assess their performance and guide future improvements in this domain:

\begin{itemize}[leftmargin=*]
    \item \textbf{RQ1: How well do code models understand API calls and imports?} The code models are assessed on their prediction of masked import or API call statements.
    The findings can help identify areas where code models may struggle and guide improvements in their API name learning by understanding this question.
    \item \textbf{RQ2: Do code models understand API import aliases?} 
    This RQ further assesses code models' understanding of API import aliases. 
    The quizzes are designed based on an alias-defining import statement followed by an API call based on the imported module.
    Tokens are selectively masked out for the statements.

    \item \textbf{RQ3: Will natural language context affect the API name knowledge probing results?} We investigate whether the overall performance of the models can be improved by incorporating natural language queries. The findings can guide the development of techniques that leverage contextual information to enhance code models' API name learning.
    \item \textbf{RQ4: How well can code models memorize and generalize on API names?} We divide APIs into two groups based on whether they are seen during the training phase.
    The outcomes indicate if the code models possess robust generalization capabilities and appropriate memorization skills. Models' ability to apply learned knowledge to unseen APIs would be helpful for API namespace design.

\end{itemize}

For the evaluation, we construct the first benchmark on Python API name knowledge, \texttt{PyINK}, and analyze 10 pre-trained code models, including the variants of CodeBERT~\cite{feng2020codebert}, GraphCodeBERT~\cite{guo2021graphcodebert}, PLBART~\cite{ahmad2021unified}. 
Our work is complementary to developing more advanced techniques for modeling API representations in code, thereby enhancing the efficiency and accuracy of code models. Additionally, the insights gained from this work can pave the way for a deeper understanding of how pre-training impacts the performance of code models, facilitating more informed design decisions in this domain.
In summary, our main contributions include:
\begin{itemize}
    \item A cloze-style evaluation framework \texttt{INK}, to probe and benchmark the knowledge of API names in pre-trained code models.
    \item An implementation of \texttt{INK}, which lowers the bar for designing probing techniques on API name knowledge.
    \item An evaluation dataset based on \texttt{INK}, \texttt{PyINK}, containing diverse API name cloze-style pop quizzes.
    \item A comprehensive study on understanding the Python API name knowledge of pre-trained code models via \texttt{PyINK}.
\end{itemize}
 
\section{Background and Related Work}
This section provides a comprehensive overview of the background and related work that form the foundation of our research. Firstly, we introduce three prominent families of pre-trained code models, namely CodeBERT~\cite{feng2020codebert}, GraphCodeBERT~\cite{guo2021graphcodebert}, and PLBART~\cite{ahmad2021unified}, which have been widely adopted in recent studies. Besides, we present a review on knowledge probing in language models, which has emerged as a crucial research area for enhancing the interpretability and understanding of such models. Finally, we discuss how existing works on deep API learning are different from our study. Through these discussions, we establish the contextual and theoretical framework necessary to fully appreciate the novelty and significance of our proposed approach.
\subsection{Pre-trained Code Models}
Pre-trained language models, such as BERT~\cite{devlin2019bert}, are commonly utilized for knowledge transfer in various downstream natural language processing tasks~\cite{qiu2020pre}. These models are trained on extensive NL corpora and fine-tuned on small labeled datasets for different tasks. They capture contextual linguistic information and eliminate the necessity for task-specific models. Similarly, pre-trained code models, such as CodeBERT, GraphCodeBERT, and PLBART, have been developed to leverage the ``naturalness'' of software~\cite{hindle2016naturalness}. These models excel in various code-related downstream tasks, such as code completion, code-to-text summarization, and code-to-code translation.

\subsubsection{CodeBERT} 
CodeBERT is a sophisticated multilingual pre-trained code model built on the BERT architecture. It excels in understanding the semantic connections between natural language (NL) and programming language (PL) through masked language modeling and replaced token detection. CodeBERT achieves state-of-the-art results in NL code search and code documentation generation tasks, making it a valuable tool for academic writing and review.

Pre-trained on the CSNet dataset~\cite{husain2019codesearchnet}, CodeBERT encompasses a diverse range of NL and PL instances, including code snippets, comments, and unimodal code. Its impressive performance is achieved without relying on explicit indicators to differentiate PLs. By fine-tuning its parameters, CodeBERT consistently demonstrates exceptional proficiency, reinforcing its position as a top choice for academic writers and reviewers.

\subsubsection{GraphCodeBERT} 

GraphCodeBERT is a pre-trained code model that revolutionizes code understanding by considering the inherent structure of code. Unlike other models that rely on complex abstract syntax trees (AST), GraphCodeBERT leverages data flow during pre-training to capture the semantic-level structure and encode variable relationships. By focusing on the "where-the-value-comes-from" aspect, GraphCodeBERT achieves more efficient and effective code representation~\cite{guo2021graphcodebert}.

Pre-trained on the same CSNet dataset as CodeBERT, GraphCodeBERT employs three distinct objectives during pre-training. It involves masked language modeling, data flow edge prediction, and variable alignment across source code and data flow. This comprehensive approach enables GraphCodeBERT to excel in four key downstream tasks: text-code search, clone detection, code translation, and code refinement, outperforming CodeBERT in all of these areas. The model's superior performance validates its efficacy and establishes GraphCodeBERT as a cutting-edge solution in the field of code analysis and understanding.

\subsubsection{PLBART} 
PLBART is a bidirectional and autoregressive code model that exhibits exceptional proficiency in performing a wide range of code summarization, generation, and translation tasks. The design of PLBART was inspired by BART~\cite{lewis2020bart}, which is a language model based on denoising autoencoding sequence-to-sequence pre-training. With a similar setup, PLBART is pre-trained using three denoising pre-training objectives, namely, token masking, token deletion, and token infilling. The first two strategies involve randomly sampling tokens and replacing them with a mask token or deleting them from the input sequence. Conversely, in token infilling, a Poisson distribution ($\lambda$ = 3.5) is used to draw the lengths of text spans that are sampled and replaced with a single mask token. In each instance, 35\% of the tokens are masked.

In contrast to CodeBERT and GraphCodeBERT, PLBART is pre-trained on a vast collection of Java and Python functions, as well as NL descriptions sourced from GitHub and StackOverflow. Based on evaluations, PLBART has demonstrated remarkable generalization capabilities and superior performance when applied to several multilingual downstream tasks.

\begin{figure*}[!ht]
    \centering
    \includegraphics[width=\textwidth]{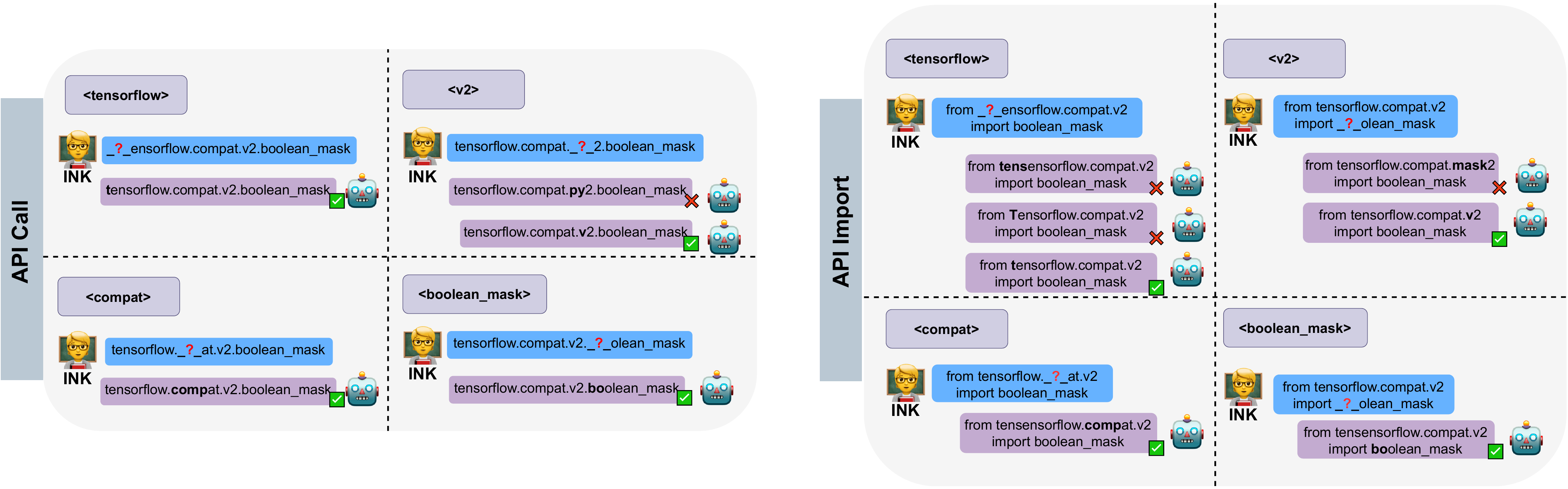}
    \caption{API fully qualified names such as \texttt{<tensorflow.compat.v2.boolean\_mask>} can be formalized into the cloze-style tests from two perspectives, API calls and API imports. The example shows the top-$k$ predictions of CodeBERT-MLM model for each test. Note that the dialogues are for illustration only.}
    \label{fig:motivation}
\end{figure*}
\subsection{Knowledge Probing}
The evaluation of internal representations and knowledge of language models is a fundamental and critical process, which involves the technique of knowledge probing~\cite{liu2023pre}. This approach entails presenting a set of questions or statements to the model to assess its understanding of specific concepts or relationships. The inputs are typically presented as cloze sentences with particular concepts or relationships masked as discrete prompts to test the model's performance. We formalize the knowledge probing approach by considering the input cloze sentence, denoted as $S$, where ``Alan Turing was born in [\texttt{MASK}]'' is an example.

Formally, we define the knowledge probing approach as follows,
\begin{align}
f(S) = \frac{1}{|S|}\sum_{k=1}^{|S|}log(P(t_k)|S;\theta), t_k \in \mathcal{V}(\theta)
\end{align}
where $\theta$ represents the model parameters, $\mathcal{V}(\theta)$ denotes the vocabulary learned by the language model, and $t_k$ is a token inside the model's vocabulary $\mathcal{V}(\theta)$. The contextualized likelihood $f(S)$ represents the possibility of replacing [\texttt{MASK}] with the token $t_k$ as per the model's prediction. The final prediction corresponds to the token $t_k$ that maximizes $f(S)$.

Knowledge probing is an essential technique for identifying areas where the model requires improvement, understanding how the model processes and represents information, and exploring the underlying knowledge. Furthermore, knowledge probing enables the development of more robust and reliable language models that reflect human understanding of language and the world.

Factual probing is an early application of prompting methods in natural language processing, with the goal of quantifying the factual knowledge encoded in pre-trained language models. This task involves transforming the input into a cloze prompt, either manually crafted or automatically discovered, to retrieve knowledge. Relevant datasets such as LAMA~\cite{petroni2019language} and X-FACTR~\cite{jiang2020x} have been established for evaluating models in fact retrieval. Researchers have explored discrete template search~\cite{petroni2020context,petroni2019language,perez2021true} and continuous template learning, as well as prompt ensemble learning~\cite{qin2021learning,jiang2020can}, to improve the effectiveness of factual probing. The existing body of research demonstrates that pre-trained language models contain substantial factual knowledge, which can be efficiently accessed using various prompting methods.

\subsection{Generation-based API Recommendation}
API recommendation is a challenging task that involves providing a specific API based on an NL query. Previous research has focused on two approaches: (1) Rank-based recommendation and (2) Generation-based recommendation. Rank-based API recommendation utilizes a knowledge graph or knowledge base to search for the most suitable APIs based on the semantic meaning of the NL query~\cite{huang2018api,rahman2016rack,thung2013automatic}. As these methods do not require learning, we discuss another line of work where deep-learning-based approaches are used for API recommendation to generate API sequences using NL queries, close to the scope of our work.

DeepAPI~\cite{gu2016deep} was the first to tackle this task by formulating it as a machine translation problem and employing an end-to-end supervised learning model. Subsequently, researchers have investigated the effectiveness of fine-tuning pre-trained code models to perform the API learning task~\cite{martin2022deep,hadi2022effectiveness}. While these studies demonstrate that code models achieve better performance on API learning, they suffer from four main drawbacks: (1) Code models are only fine-tuned with a few APIs and can not be generalized to generate APIs in the wild. (2) Fine-tuning code models to synthesize API sequences lacks interpretability for capturing API knowledge during pre-training. (3) The API learning task is based on NL inputs that only determine the semantic understanding of API sequence usage. (4) Evaluation of API learning is solely based on BLEU score~\cite{papineni2002bleu}, which measures the similarity between synthesized API sequences and references and fails to reflect synthesis correctness.

\section{\texttt{INK}: An evaluation framework of AP\underline{I} \underline{N}ame \underline{K}nowledge}

\begin{figure*}
    \centering
    \includegraphics[width=0.9\textwidth]{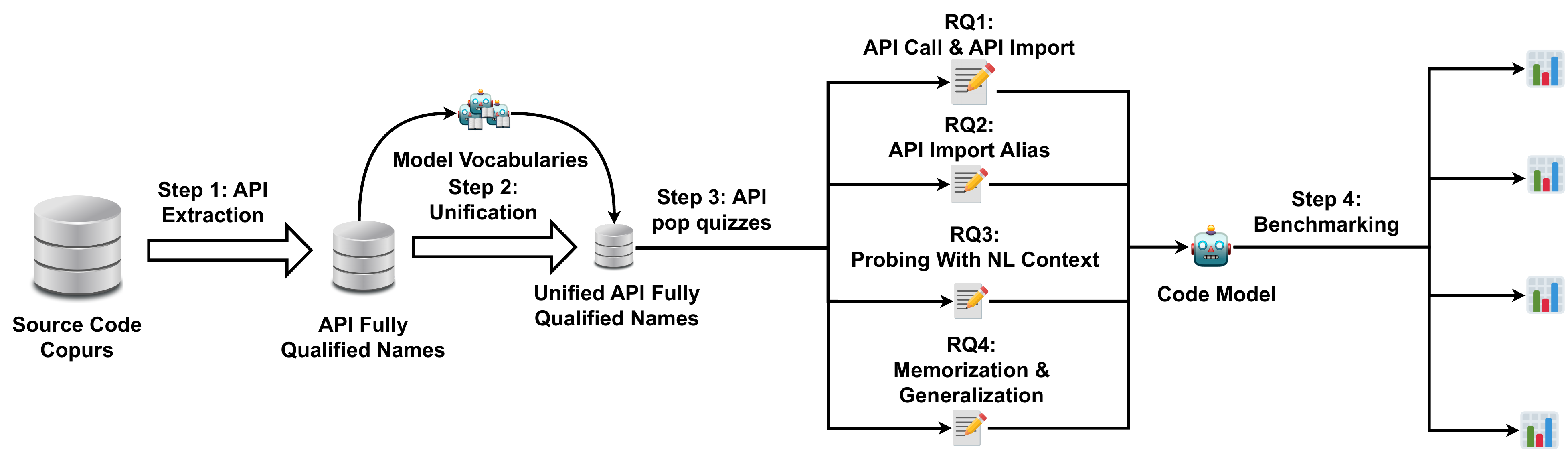}
    \caption{An overview of the \texttt{INK} framework.}
    \label{fig:pipeline}
\end{figure*}
\subsection{Motivation}
\label{sec:3}
Previous research in natural language processing has utilized cloze sentences for token prediction as a means of interpreting the knowledge encoded by pre-trained language models. Building upon this work, we examine probing API name knowledge in CodeBERT-MLM - a variant of CodeBERT pre-trained solely on mask language modeling - with cloze-style sentences serving as pop quizzes, as depicted in Figure~\ref{fig:motivation}. We use \texttt{<tensorflow.compat.v2.boolean\\\_mask>} as an example and transform it into a cloze-style pop quiz, as shown in Figure~\ref{fig:motivation}. In this study, we define the API module levels as each hierarchical level within the fully qualified name, separated by a period. There are four module levels in the API call statement: (1) \texttt{<tensorflow>} representing the top module level, (2) \texttt{<compat>} as the second module level, (3) \texttt{<v2>} as the third module level, and (4) \texttt{<boolean\_mask>} as the last call level. For each level, we request CodeBERT-MLM to fill in the blank via \textbf{first} token prediction, as determined by its tokenizer. As our results demonstrate, CodeBERT-MLM correctly predicts the masked token on the first attempt, except for the third level of \texttt{<v2>}. We contend that code models, such as CodeBERT-MLM, can learn API names during function pre-training.

Given that \texttt{<tensorflow.compat.v2.boolean\\\_mask>} can be reconstituted to the form of an API import statement, we then investigate how well CodeBERT-MLM understands the API import statements. As demonstrated in Figure~\ref{fig:motivation}, we transform the API import statement into four cloze templates by masking the first token of each module level, similar to the ones of the API call. From the results illustrated in Figure~\ref{fig:motivation}, we discover that predicting some API modules at the first shot is challenging for CodeBERT-MLM, unlike the case of API call probing. This behavior suggests that code models possess varying degrees of knowledge in API import and API call statements.

Having identified several potential patterns from our preliminary knowledge probing study, which provides clues to API name knowledge, we find it necessary to further explore this phenomenon through quantitative analysis and systematic evaluation. Motivated by the aforementioned observations, this paper investigates whether pre-trained code models learn the API names and to what extent they store API name knowledge, by conducting knowledge probing as the pop quiz. Specifically, we analyze two perspectives of API names, i.e., API call and API import, within the purview of code models.

\subsection{API-to-Pop-Quiz Conversion}

We consider three main types of transformation of evaluating API name knowledge: \textbf{API calls}, \textbf{API imports} and \textbf{API import aliases}. As aforementioned in Section~\ref{sec:3}, cloze-style pop quizzes are structured based on each call level split by the delimiters of ``.'', ``from'' and ``import''. To benchmark models fairly, we construct pop quizzes by unifying the entire vocabulary of each model. For all the evaluations, we follow previous work of knowledge probing in language models~\cite{petroni2019language} and choose to evaluate the prediction of a single token masking in the pop quiz. We provide the detailed design of each process as follows.

\subsubsection{\textbf{Evaluation Design on API Call}}
\hfill\\
We treat each API call as a modular pattern on the basis of each module level. To formalize the pop quiz construction on API calls, an example of API call \texttt{<A.B.C>} and a code model $\mathcal{M}$ are given to demonstrate the workflow. The model $\mathcal{M}$ firstly tokenizes the API as follows,
\begin{gather*}
    \mathcal{M}(\texttt{<A.B.C>}) \rightarrow\\
    \bigg\{\{t_1^A, t_2^A, \dots, t_{N_a}^A\},t^{dot},
    \{t_1^B, t_2^B, \dots, t_{N_b}^B\},t^{dot},\\
    \{t_1^C, t_2^C, \dots, t_{N_c}^C\}\bigg\}
\end{gather*}
where each $t$ represents the token produced by the model $\mathcal{M}$, and $N$ represents the length of tokens in each level. For each level, the tokens are grouped by $\{\dots\}$. When converting the tokenized API to the pop quiz, we mask a specific token by replacing $t$ with a ``\texttt{[MASK]}'' in each level. To visualize the pop quiz input, we mask the last token in the second module level of \texttt{<A.B.C>} as follows:
\begin{gather*}
    \texttt{<A.B.C>} \rightarrow \texttt{<A.B'[MASK].C>} \rightarrow \texttt{<A.B'\underline{\hspace{0.3cm}}.C>}
\end{gather*}
where $B'$ is the concatenation of $\{t_1^B, \dots, t_{N_b-1}^B\}$. We prompt the model $\mathcal{M}$ to fill in the blank of \texttt{<A.B'\underline{\hspace{0.3cm}}.C>} via mask prediction.
\subsubsection{\textbf{Evaluation Design on API Import}}
\hfill\\
We explore the evaluation design on the API import statement of ``\texttt{from} \dots \texttt{import} \dots''. Similarly, we consider the example of \texttt{<from A.B import C>}. Using the model $\mathcal{M}$ to tokenize the API import, we can devise the following tokens:
\begin{gather*}
\mathcal{M}(\texttt{<from A.B import C>}) \rightarrow\\
    \bigg\{t^{from},\{t_1^A, t_2^A, \dots, t_{N_a}^A\},t^{dot},\\
    \{t_1^B, t_2^B, \dots, t_{N_b}^B\},t^{import},
    \{t_1^C, t_2^C, \dots, t_{N_c}^C\}\bigg\}
\end{gather*}
We visualize an example of API import quiz, where the first token in the bottom level of \texttt{<from A.B import C>} is masked:
\begin{gather*}
\texttt{<from A.B import C>}\\
    \rightarrow
    \texttt{<from A.B import [MASK]C'>}\\
    \rightarrow \texttt{<from A.B import \underline{\hspace{0.3cm}}C'>}
\end{gather*}
where $C'$ is the concatenation of $\{t_2^C, \dots, t_{N_c}^C\}$. We probe the model $\mathcal{M}$ to fill in the blank of \texttt{<from A.B import \underline{\hspace{0.3cm}}C'>} via mask prediction.

\subsubsection{\textbf{Evaluation Design on API Import Alias}}
\hfill

We note that import aliases are supported in some programming languages, such as Python. For example, ``\texttt{import} \dots as \dots'' and ``\texttt{from} \dots \texttt{import} \dots as \dots'' are the typical import alias syntax. Therefore, we further examine pre-trained code model's understanding of the aliases of API calls after packages and libraries imports. We illustrate the design choice via the example of \texttt{<import A as K \textbackslash n K.B.C>}, where \texttt{<K>} is the alias and \texttt{K.B.C} is the API call statement. After being tokenized by the model $\mathcal{M}$, the example is formalized as follows:
\begin{gather*}
\mathcal{M}(\texttt{<import A as K \textbackslash n K.B.C>}) \rightarrow\\
    \bigg\{t^{import},\{t_1^A, t_2^A, \dots, t_{N_a}^A\},t^{as},
    \{t_1^K, t_2^K, \dots, t_{N_a}^K\},t^{newline},\\
    \{t_1^K, t_2^K, \dots, t_{N_a}^K\},t^{dot},\\
    \{t_1^B, t_2^B, \dots, t_{N_b}^B\},t^{dot},
    \{t_1^C, t_2^C, \dots, t_{N_c}^C\}\bigg\}
\end{gather*}

We then convert the example to the following API alias pop quiz with the masked last token of \texttt{<C>}:
\begin{gather*}
    \texttt{<import A as K \textbackslash n K.B.C>}\\
    \rightarrow
    \texttt{<import A as K \textbackslash n K.B.C'[MASK]>}\\
    \rightarrow \texttt{<import A as K \textbackslash n K.B.C'\underline{\hspace{0.3cm}}>}
\end{gather*}
where $C'$ is the concatenation of $\{t_1^C, \dots, t_{N_c-1}^C\}$. We probe the model $\mathcal{M}$ to fill in the blank of \texttt{<import A as K \textbackslash n K.B.C'\underline{\hspace{0.3cm}}>} via mask prediction. Note that we only mask the tokens in the part of the API call after the alias name.

\subsubsection{\textbf{Evaluation of Code Models: A Unified Vocabulary Approach}}
\hfill\\
Code models may tokenize the same API call or import statement differently due to variations in their respective vocabularies. For example, an API call such as \texttt{<A.B.C>} may be tokenized as a single token by one model $\mathcal{M}_A$ with vocabulary $\mathcal{V}_A$, while another model $\mathcal{M}_B$ with vocabulary $\mathcal{V}_B$ may tokenize it into multiple tokens. To ensure a fair comparison, we generate models over a unified vocabulary that is the intersection of the vocabularies of all considered code models. We define the unified vocabulary as the set of tokens that are presented in the vocabularies of all considered code models.

To evaluate the performance of code models on this unified vocabulary, we categorize the evaluation of each name level into two types: (1) \textit{partial module masking} and (2) \textit{full module masking}. For the former, we tokenize the API call and import statements into multiple tokens and mask only one of the tokens. In contrast, when code models segment the entire module level as a single token, we denote it as the full module masking.

\subsubsection{\textbf{Benchmarking Code Models via Knowledge Probing}}
\hfill\\
We utilize the knowledge probing technique to effectively evaluate and assess the performance of code models. The primary objective of this approach is to delve into the code models' comprehension of API names and their proficiency in generating precise and contextually fitting tokens within those names. To achieve this, we present cloze-style quizzes where certain tokens are intentionally masked, prompting the code model to predict the tokens for the masked positions. To gauge the accuracy of these predictions, we compare them against the ground-truth tokens in the API names. This benchmarking process allows us to determine the correctness of predictions.
\section{Experiment Setup}
We introduce the research questions, the basic experimental setup about the datasets and models, and the evaluation metrics used throughout the evaluation. The research questions we aim to answer include:

\begin{itemize}[leftmargin=*]
  \item[] \textbf{RQ1:} How well do code models understand API calls and imports?
  \item[] \textbf{RQ2:} Do code models understand API import aliases?
  \item[] \textbf{RQ3:} Will natural language context affect the API name knowledge probing results?
  \item[] \textbf{RQ4:} How well can code models memorize and generalize on API names?
\end{itemize} 
\subsection{\texttt{PyINK}: Evaluation on Python API Name Knowledge}

\begin{table}[!h]
    \centering
    \resizebox{\columnwidth}{!}{
    \begin{tabular}{|c|c|c|c|c|}\hline
      & \textbf{First Token} & \textbf{Last Token} & \textbf{Full} & \textbf{Total}\\\hline
      \multirow{1}{*}{API Call}
      & 108,863 & 111,373 & 69,349 & 289,585 \\
      \multirow{1}{*}{API Import}
      & 108,863 & 111,373 & 69,349 & 289,585 \\
      \multirow{1}{*}{API Import Alias}
      & 7,385 &  7,122 & 3,464 & 17,971
      \\\hline
    \end{tabular}}
    \caption{Overview of the number of pop quizzes in \texttt{PyINK}.}
    \label{tab:pyink}
\end{table}
We adopt the widely-used pre-training corpus, CSNet~\cite{husain2019codesearchnet}, which is a collection of datasets and benchmarks for semantic code retrieval, containing functions and their corresponding comments of six programming languages extracted from GitHub repositories, to evaluate the API name knowledge of code models. We focus on the Python set, which contains 457,461 functions. To extract Python APIs from CSNet, we use our proposed \texttt{INK} framework, which leverages static analysis on the entire file to extract API usages, following the method described in \cite{wang2021restoring}. To achieve this, we clone all repositories corresponding to the functions in CSNet and analyze the API usage in each function. This process results in a new benchmark, denoted as \texttt{PyINK}, which is designed to evaluate the name knowledge of Python APIs. \texttt{PyINK} contains 597,141 main pop quizzes for evaluation, and we only consider \textbf{full} module masking, or partial masking of the \textbf{first} and \textbf{last} tokens of each module level to maintain consistency.

Our approach successfully extracts 79,754 unique APIs from 8,294 Python libraries in 13,519 repositories, indicating the diverse usage of Python APIs. Moreover, to ensure a fair comparison, we unify the vocabularies in the considered code models, resulting in 289,585, 289,585, and 17,971 samples for API call, import, and import with aliases statements, respectively, in the benchmark.

\subsection{Code Models}
We conduct extensive studies on ten selected models from the variants of CodeBERT, GraphCodeBERT and PLBART. We also include GPT-3.5-turbo for the assessment, though it is not specifically trained with mask prediction. We instruct GPT-3.5-turbo to predict top-20 masked tokens for each test of 1,000 randomly selected samples on API calls and imports, respectively. To guide GPT-3.5-turbo to correctly perform the prediction task, we prompt it with the instruction of ``Predict top-20 answers of \texttt{<mask>} part of the following Python API fully qualified name, pay attention to the connected characters right after \texttt{<mask>}. Note the number of masked characters is at least one. Print each of 20 answers per line with index.''. The description of each model is summarized in Table~\ref{tab:code_model}.

\begin{table}[!h]
    \centering
    \resizebox{\columnwidth}{!}{
    \begin{tabular}{|cc|cccc|}\hline
       \multicolumn{2}{|c|}{\textbf{Model}} & \textbf{pre-trained Dataset} & \textbf{Objective} & \textbf{\#Param} & \textbf{Fine-tuned}\\\hline
       \multirow{2}{*}{CodeBERT} & MLM & CSNet & MLM & 125M & \xmark\\
       & MLM-Python  & CSNet+CodeParrot* & MLM & 125M & \xmark\\\hline
       GraphCodeBERT & MLM & CSNet & MLM*  & 125M & \xmark\\\hline
        \multirow{6}{*}{PLBART}  & Base & N/A & DAE & 140M & \xmark\\
        & Large  & N/A & DAE & 406M & \xmark\\
        & CSNet & CSNet & DAE & 140M & \xmark\\\cdashline{2-6}
        & \textcolor{gray}{Sum} & \textcolor{gray}{N/A} & \textcolor{gray}{DAE} & \textcolor{gray}{140M} & \textcolor{gray}{\cmark}\\
        & \textcolor{gray}{Gen} & \textcolor{gray}{N/A}  & \textcolor{gray}{DAE} & \textcolor{gray}{140M} & \textcolor{gray}{\cmark}\\
        & \textcolor{gray}{MT} & \textcolor{gray}{N/A} & \textcolor{gray}{DAE} & \textcolor{gray}{140M} & \textcolor{gray}{\cmark}\\\hline
        GPT-3.5 & turbo &N/A & N/A & 154B & \cmark\\\hline
    \end{tabular}}
    \caption{The overview of selected code models for evaluation. Note that only the train split of CSNet is used during pre-training. MLM: Masked Language Modeling. DAE: Denoising Auto-Encoding. CodeBERT-MLM-Python uses the Python split of CodeParrot~\cite{wolf2020transformers} for continuous pre-training. \textcolor{gray}{PLBART-Sum} and \textcolor{gray}{PLBART-Gen} are the PLBART-Base model fine-tuned on CSNet Python code summarization and code generation tasks, respectively. \textcolor{gray}{PLBART-MT} adopts multitask learning and is fine-tuned on both tasks that the previous two models use.}
    \label{tab:code_model}
\end{table}

\subsection{Evaluation Metric}
We present an evaluation methodology based on rank-based metrics in the context of API name prediction. Our approach involves computing results per test sample and means across pop quizzes, utilizing the mean precision at k ($P@k$) metric. Specifically, $P@k$ is computed as 1 if the target object is ranked among the top k results, and 0 otherwise.

\section{Results}




\subsection{RQ1: How well do code models understand API calls and imports?}

\begin{table}[!h]
    \centering
    \resizebox{\columnwidth}{!}{
    \begin{tabular}{|l|c|c|c|c|c|c|c|c|c|}\hline
    \multicolumn{3}{|c|}{} & \multicolumn{1}{c|}{$P@1 \uparrow$} & \multicolumn{1}{c|}{$P@5 \uparrow$} & \multicolumn{1}{c|}{$P@10 \uparrow$}
    & \multicolumn{1}{c|}{$P@20 \uparrow$} & \multicolumn{1}{c|}{$P@30 \uparrow$}
    & \multicolumn{1}{c|}{$P@40 \uparrow$} & \multicolumn{1}{c|}{$P@50 \uparrow$}\\\hline
  \multirow{9}{*}{\rotatebox[origin=c]{90}{API Call}}
   &\multirow{2}{*}{CodeBERT} &MLM
& 23.90
& 42.34
& 50.28
& 58.17
& 62.75
& 65.87
& 68.28
    \\
    & &{MLM-Python}
& \textbf{29.35}
& \textbf{47.51}
& \textbf{55.67}
& \textbf{63.61}
& \textbf{68.06}
& \textbf{71.06}
& \textbf{73.33}

    \\\cline{2-10}
    & GraphCodeBERT & MLM
& 25.89
& 43.30
& 50.74
& 58.28
& 62.73
& 65.87
& 68.26

    \\\cline{2-10}
    & \multirow{6}{*}{PLBART} & Base
& 1.65
& 6.58
& 10.43
& 15.71
& 19.40
& 22.44
& 24.99

    \\
    & & {Large}
& 2.29
& 8.63
& 13.22
& 18.62
& 22.19
& 25.02
& 27.40
    \\
    & &CSNet
& 2.57
& 10.78
& 15.57
& 21.49
& 25.55
& 28.76
& 31.41
    \\\cdashline{3-10}
    & &\textcolor{gray}{Sum}
& 2.11
& 9.65
& 14.46
& 19.98
& 23.53
& 26.21
& 28.46
    \\
    & &\textcolor{gray}{Gen}
& 4.31
& 14.51
& 20.69
& 27.21
& 31.27
& 34.23
& 36.59
    \\
    & &{\textcolor{gray}{MT}}
& 4.79
& 15.80
& 22.71
& 30.27
& 34.83
& 38.10
& 40.67
\\\cline{2-10}
    & \multirow{1}{*}{GPT-3.5} & turbo*
&16.00
&34.10
&39.80
&44.20
&-
&-
&-
    \\
    \hline\hline

    \multirow{9}{*}{\rotatebox[origin=c]{90}{API Import}}
   &\multirow{2}{*}{CodeBERT} &MLM
& 26.33
& 43.87
& 51.19
& 58.63
& 62.89
& 65.85
& 68.16
    \\
    & &{MLM-Python}
& 29.67
& \textbf{49.26}
& \textbf{56.82}
& \textbf{64.18}
& \textbf{68.34}
& \textbf{71.23}
& \textbf{73.39}

    \\\cline{2-10}
    & GraphCodeBERT & MLM
& \textbf{30.76}
& 48.24
& 55.35
& 62.50
& 66.68
& 69.55
& 71.73

    \\\cline{2-10}
    & \multirow{6}{*}{PLBART} & Base
& 2.64
& 12.96
& 20.12
& 28.21
& 33.22
& 36.90
& 39.83
    \\
    & & {Large}
& 5.66
& 18.03
& 25.72
& 34.20
& 39.18
& 42.69
& 45.44
    \\
    & &CSNet
& 3.44
& 14.90
& 21.95
& 30.31
& 35.79
& 39.86
& 43.11
    \\\cdashline{3-10}
    & &\textcolor{gray}{Sum}
& 2.88
& 12.26
& 18.90
& 26.09
& 30.44
& 33.61
& 36.18
    \\
    & &\textcolor{gray}{Gen}
& 3.83
& 15.25
& 23.59
& 32.52
& 37.72
& 41.38
& 44.25
    \\
    & &{\textcolor{gray}{MT}}
& 4.92
& 19.16
& 28.02
& 36.65
& 41.65
& 45.33
& 48.16
    \\\cline{2-10}
    & \multirow{1}{*}{GPT-3.5} & turbo*
&20.30
&37.80
&45.20
&50.50
&-
&-
&-\\\hline
    \end{tabular}
    }
    \caption{$P@k$ scores on selected code models, focusing API calls and API imports.}
    \label{tab:rq2_table}
\end{table}

\begin{table*} []
    \centering
    \resizebox{\textwidth}{!}{
    \begin{tabular}{lllll}\hline
        & \textbf{Test} & \textbf{Answer} & \textbf{CodeBERT-MLM-Python} & \textbf{GraphCodeBERT-MLM} \\\hline
        \multirow{5}{*}{\rotatebox[origin=c]{90}{API Call}}
&
stsci.tools.fileutil.buildFITS\underline{\hspace{0.5cm}}(
&
Name
&
\colorbox{mygray}{file}[0.10], \colorbox{mygray}{File}[0.06], \colorbox{mygray}{Image}[0.04], \colorbox{mygray}{image}[0.03], \colorbox{mygray}{Data}[0.02]
&
\colorbox{mygray}{)\textbackslash}[0.42], \colorbox{mygray}{):}[0.08], \colorbox{mygray}{()}[0.06], \colorbox{mygray}{);}[0.05], \colorbox{mygray}{ )}[0.04]
\\
&
win32\underline{\hspace{0.5cm}}.QueryServiceStatusEx(
&
service
&
\colorbox{mygray}{api}[0.59], \colorbox{mygray}{gui}[0.25], \colorbox{pink}{service}[0.04], \colorbox{mygray}{security}[0.02], \colorbox{mygray}{com}[0.01]
&
\colorbox{pink}{service}[0.33], \colorbox{mygray}{Service}[0.09], \colorbox{mygray}{net}[0.08], \colorbox{mygray}{api}[0.06], \colorbox{mygray}{API}[0.05]
\\
&
demand\underline{\hspace{0.5cm}}.bdew.ElecSlp(
&
lib
&
\colorbox{mygray}{2}[0.12], \colorbox{pink}{lib}[0.09], \colorbox{mygray}{1}[0.07], \colorbox{mygray}{3}[0.03], \colorbox{mygray}{py}[0.02]
&
\colorbox{mygray}{ com}[0.47], \colorbox{mygray}{ org}[0.17], \colorbox{mygray}{ google}[0.16], \colorbox{mygray}{com}[0.03], \colorbox{mygray}{org}[0.02]
\\
&
cltk.prosody.latin.string\_\underline{\hspace{0.5cm}}.move\_consonant\_right(
&
utils
&
\colorbox{mygray}{left}[0.02], \colorbox{mygray}{table}[0.02], \colorbox{mygray}{translation}[0.02], \colorbox{mygray}{char}[0.02], \colorbox{mygray}{right}[0.02]
&
\colorbox{pink}{utils}[0.28], \colorbox{mygray}{list}[0.16], \colorbox{mygray}{selection}[0.10], \colorbox{mygray}{table}[0.08], \colorbox{mygray}{util}[0.04]
\\
&
\underline{\hspace{0.5cm}}fuzz.rand.randint(
&
gram
&
\colorbox{mygray}{\textbackslash n}[0.40], \colorbox{mygray}{\#}[0.34], \colorbox{mygray}{py}[0.09], \colorbox{mygray}{\#\#}[0.02], \colorbox{mygray}{.}[0.01]
&
\colorbox{mygray}{\#}[0.26], \colorbox{mygray}{\^{}}[0.05], \colorbox{mygray}{open}[0.05], \colorbox{mygray}{!}[0.03], \colorbox{mygray}{test}[0.03]
\\\hline
        \multirow{5}{*}{\rotatebox[origin=c]{90}{API Import}}
&
from stsci.tools.fileutil import buildFITS\underline{\hspace{0.5cm}}
&
Name
&
\colorbox{mygray}{\textbackslash n}[0.94], \colorbox{mygray}{\textbackslash n\textbackslash n}[0.03], \colorbox{mygray}{File}[0.00], \colorbox{mygray}{file}[0.00], \colorbox{mygray}{\_}[0.00]
&
\colorbox{mygray}{File}[0.38], \colorbox{mygray}{file}[0.23], \colorbox{mygray}{Filename}[0.02], \colorbox{mygray}{Tools}[0.02], \colorbox{mygray}{Files}[0.01]
\\
&
from win32\underline{\hspace{0.5cm}} import QueryServiceStatusEx
&
service
&
\colorbox{mygray}{api}[0.43], \colorbox{pink}{service}[0.07], \colorbox{mygray}{query}[0.07], \colorbox{mygray}{security}[0.06], \colorbox{mygray}{db}[0.04]
&
\colorbox{pink}{service}[0.35], \colorbox{mygray}{file}[0.22], \colorbox{mygray}{net}[0.19], \colorbox{mygray}{db}[0.02], \colorbox{mygray}{security}[0.02]
\\
&
from demand\underline{\hspace{0.5cm}}.bdew import ElecSlp
&
lib
&
\colorbox{mygray}{er}[0.03], \colorbox{mygray}{1}[0.03], \colorbox{mygray}{y}[0.02], \colorbox{mygray}{en}[0.02], \colorbox{mygray}{2}[0.02]
&
\colorbox{mygray}{igo}[0.05], \colorbox{mygray}{y}[0.02], \colorbox{mygray}{ com}[0.02], \colorbox{mygray}{en}[0.02], \colorbox{mygray}{ache}[0.02]
\\
&
from cltk.prosody.latin.string\_\underline{\hspace{0.5cm}} import move\_consonant\_right
&
utils
&
\colorbox{mygray}{move}[0.14], \colorbox{mygray}{right}[0.08], \colorbox{mygray}{tools}[0.04], \colorbox{mygray}{translation}[0.03], \colorbox{mygray}{left}[0.03]
&
\colorbox{pink}{utils}[0.56], \colorbox{mygray}{selection}[0.10], \colorbox{mygray}{util}[0.06], \colorbox{mygray}{list}[0.03], \colorbox{mygray}{table}[0.02]
\\
&
from \underline{\hspace{0.5cm}}fuzz.rand import randint
&
gram
&
\colorbox{mygray}{ py}[0.29], \colorbox{mygray}{ sk}[0.06], \colorbox{mygray}{ fuzzy}[0.05], \colorbox{mygray}{ core}[0.03], \colorbox{mygray}{ ham}[0.03]
&
\colorbox{mygray}{ ham}[0.43], \colorbox{mygray}{ exp}[0.12], \colorbox{mygray}{ math}[0.05], \colorbox{mygray}{ http}[0.04], \colorbox{mygray}{ easy}[0.03]
\\\hline
   \end{tabular}}
    \caption{Examples of prediction for CodeBERT-MLM-Python and GraphCodeBERT-MLM on API calls and imports. The last two columns reports the top five tokens generated together with the associated probabilities (in square brackets).}
    \label{tab:rq2_example}
\end{table*}

\begin{table}[!h]
    \centering
    \resizebox{\columnwidth}{!}{
    \begin{tabular}{|l|cc|c|c|c|c|c|c|c|c|}\hline
    \multicolumn{4}{|c|}{} 
    & \multicolumn{1}{c|}{$P@1 \uparrow$} 
    & \multicolumn{1}{c|}{$P@5 \uparrow$}
    & \multicolumn{1}{c|}{$P@10 \uparrow$} 
    & \multicolumn{1}{c|}{$P@20 \uparrow$}
    & \multicolumn{1}{c|}{$P@30 \uparrow$} 
    & \multicolumn{1}{c|}{$P@40 \uparrow$}
    & \multicolumn{1}{c|}{$P@50 \uparrow$}\\\cline{1-11}

    \multirow{20}{*}{\rotatebox[origin=c]{90}{API Call With Alias}}
&\multirow{4}{*}{CodeBERT} &\multirow{2}{*}{MLM}
& Alias
& 15.83
& 32.54
& 40.40
& 48.46
& 53.50
& 56.93
& 59.88
\\
& & &Adv. Alias
& 14.10
& 29.84
& 37.66
& 45.89
& 50.86
& 54.37
& 57.16
    \\\cline{4-11}
    & &\multirow{2}{*}{MLM-Python}
& Alias
& \textbf{24.17}
& \textbf{42.30}
& \textbf{50.19}
& \textbf{58.02}
& \textbf{62.68}
& \textbf{66.05}
& \textbf{68.37}
\\
& & &Adv. Alias
& \textcolor{gray}{\uwave{20.35}}
& \textcolor{gray}{\uwave{37.30}}
& \textcolor{gray}{\uwave{45.33}}
& \textcolor{gray}{\uwave{53.42}}
& \textcolor{gray}{\uwave{58.22}}
& \textcolor{gray}{\uwave{61.72}}
& \textcolor{gray}{\uwave{64.37}}

    \\\cline{2-11}
    & \multirow{2}{*}{GraphCodeBERT} & \multirow{2}{*}{MLM}
& Alias
& 18.91
& 36.85
& 43.99
& 51.66
& 56.18
& 59.58
& 62.17
\\
& & &Adv. Alias
& 16.89
& 33.63
& 40.88
& 48.48
& 53.12
& 56.44
& 59.09
\\\cline{2-11}

    & \multirow{12}{*}{PLBART} & \multirow{2}{*}{Base}
& Alias
& 0.84
& 4.20
& 8.05
& 14.56
& 19.34
& 23.46
& 26.84
\\
& & &Adv. Alias
& 0.38
& 2.62
& 5.60
& 10.81
& 14.99
& 18.45
& 21.47
    \\\cline{4-11}
    & & \multirow{2}{*}{Large}
& Alias
& 3.36
& 11.64
& 18.34
& 26.15
& 30.88
& 34.71
& 37.58
\\
& & &Adv. Alias
& 1.49
& 7.19
& 12.70
& 19.63
& 23.92
& 27.08
& 29.69
    \\\cline{4-11}
    & &\multirow{2}{*}{CSNet}
& Alias
& 1.75
& 7.86
& 12.37
& 18.59
& 22.93
& 26.87
& 30.56
\\
& & &Adv. Alias
& 1.68
& 6.63
& 10.39
& 15.48
& 19.30
& 22.45
& 25.21
    \\\cdashline{3-3}\cline{4-11}
    & &\multirow{2}{*}{\textcolor{gray}{Sum}}
& Alias
& 1.21
& 4.93
& 8.44
& 14.20
& 18.14
& 21.50
& 24.21
\\
& & &Adv. Alias
& 1.02
& 3.75
& 6.36
& 10.43
& 13.74
& 16.55
& 18.95
    \\\cline{4-11}
    & &\multirow{2}{*}{\textcolor{gray}{Gen}}
& Alias
& 3.43
& 12.87
& 19.29
& 26.85
& 31.66
& 35.05
& 37.61
\\
& & &Adv. Alias
& 2.13
& 8.54
& 13.55
& 19.79
& 23.89
& 27.08
& 29.67
    \\\cline{4-11}
    & &\multirow{2}{*}{\textcolor{gray}{MT}}
& Alias
& 3.56
& 11.68
& 18.61
& 26.42
& 31.35
& 34.80
& 37.82
\\
& & &Adv. Alias
& 2.39
& 8.08
& 12.98
& 19.72
& 24.47
& 28.03
& 30.94
\\\cline{2-11}
    & \multirow{2}{*}{GPT-3.5} & \multirow{2}{*}{turbo*}
& Alias
&26.80
&45.20
&49.80
&54.90
&-
&-
&-
\\
& & &Adv. Alias
&24.30
&42.60
&47.20
&51.45
&-
&-
&-
    \\\hline
    \end{tabular}}
    \caption{Comparison of $P@k$ scores on API import alias quizzes among selected code models.}
    \label{tab:rq_alias}
\end{table}
Our evaluation assesses the capability of pre-trained code models to encode knowledge of Python API names for both API calls and imports. We computed $P@k$ scores for each masking strategy and present the results in Table~\ref{tab:rq2_table}. In addition, we provide a few examples in Table~\ref{tab:rq2_example}. Firstly, we have observed that the relative performances of different code models remain consistent when we vary the value of $k$ in the $P@k$ metric. Secondly, as $k$ increases, the improvement in performance for each model becomes less significant. These observations provide strong evidence to support the effectiveness of the \texttt{PyINK} benchmark. When comparing the model variants, our analysis reveals that CodeBERT-MLM-Python and GraphCodeBERT-MLM demonstrate superior performance on API calls and imports compared to other models. However, their overall precision of 30\% measured by $P@1$ falls short of perfection, indicating a lack of knowledge about API names. While we expected GPT-3.5-turbo to have a better understanding of API names, it shows that the model performs slightly worse than CodeBERT-MLM. 
Additionally, our comparison shows that PLBART variants perform much worse on understanding Python API name knowledge than BERT-like models, which can be explained by the pre-training objectives of PLBART. PLBART-Large consistently outperforms other variants, indicating that model size may be an important factor in the amount of stored API name knowledge. However, this finding should be interpreted in light of the scaling law of mixed-modal language models~\cite{aghajanyan2023scaling}, which suggests that larger models are likely to achieve better performance on downstream tasks, such as code generation. Finally, we find that pre-trained data can influence the understanding of API names to some extent, as shown by the performance gap between PLBART-Base and PLBART-CSNet. Our results indicate that fine-tuning on code generation tasks can improve the performance of pre-trained models, while text generation tasks may negatively impact them.

\finding{of RQ1}{Although CodeBERT-MLM-Python and GraphCodeBERT-MLM show superior performance on API call and import name knowledge among code models, there is a significant margin for the improvement.
}

\subsection{RQ2: Do code models understand API import aliases?}
To assess the code models on the knowledge of API import aliases, we pair the 17,971 API import alias quizzes with the adversarial examples designed to test the model's robustness. To construct the adversarial set, we randomly selected 10 distinct aliases that are used in other modules and replaced the original aliases in the quizzes with these new aliases. For example, ``\texttt{import numpy as np \textbackslash n np.load\underline{\hspace{0.3cm}}}('' will be transformed to ``\texttt{import numpy as pmd \textbackslash n pmd.load\underline{\hspace{0.3cm}}}('' via the replacement of ``\texttt{np}''. In the end, we collected 179,710 adversarial quizzes.
\begin{table*}[]
    \centering
    \resizebox{\textwidth}{!}{\begin{tabular}{lllll}\hline
         & \textbf{Test} & \textbf{Answer} & \textbf{CodeBERT-MLM-Python} & \textbf{GraphCodeBERT-MLM} \\\hline
        \multirow{10}{*}{\rotatebox[origin=c]{90}{API Alias}}
&
import memote.suite.cli.callbacks as callbacks
&
\multirow{2}{*}{
git
}
&
\multirow{2}{*}{
\colorbox{mygray}{check}[0.19], \colorbox{mygray}{load}[0.16], \colorbox{mygray}{get}[0.07], \colorbox{mygray}{register}[0.07], \colorbox{mygray}{update}[0.04]
}
&
\multirow{2}{*}{
\colorbox{mygray}{on}[0.22], \colorbox{mygray}{set}[0.09], \colorbox{mygray}{register}[0.05], \colorbox{mygray}{command}[0.03], \colorbox{mygray}{server}[0.03]
}
\\& 
callbacks.<mask>\_installed(
&&&
\\\cdashline{2-5}
&
import weka.flow.conversion as conversion
&
\multirow{2}{*}{
Pass
}
&
\multirow{2}{*}{
\colorbox{mygray}{pass}[0.50], \colorbox{pink}{Pass}[0.17], \colorbox{mygray}{run}[0.05], \colorbox{mygray}{Go}[0.02], \colorbox{mygray}{Run}[0.01]
}
&
\multirow{2}{*}{
\colorbox{mygray}{pass}[0.66], \colorbox{pink}{Pass}[0.15], \colorbox{mygray}{cycle}[0.08], \colorbox{mygray}{write}[0.04], \colorbox{mygray}{walk}[0.01]
}
\\& 
conversion.<mask>Through(
&&&
\\\cdashline{2-5}
&
import memote.support.helpers as helpers
&
\multirow{2}{*}{
find
}
&
\multirow{2}{*}{
\colorbox{mygray}{test}[0.29], \colorbox{mygray}{get}[0.10], \colorbox{mygray}{load}[0.08], \colorbox{mygray}{check}[0.07], \colorbox{mygray}{create}[0.05]
}
&
\multirow{2}{*}{
\colorbox{mygray}{start}[0.27], \colorbox{mygray}{allow}[0.14], \colorbox{mygray}{stop}[0.08], \colorbox{mygray}{try}[0.04], \colorbox{mygray}{begin}[0.04]
}
\\& 
helpers.<mask>\_converting\_reactions(
&&&
\\\cdashline{2-5}
&
import pandas as pd
&
\multirow{2}{*}{
Store
}
&
\multirow{2}{*}{
\colorbox{pink}{Store}[0.35], \colorbox{mygray}{5}[0.30], \colorbox{mygray}{S}[0.11], \colorbox{mygray}{Parser}[0.04], \colorbox{mygray}{4}[0.04]
}
&
\multirow{2}{*}{
\colorbox{mygray}{5}[0.34], \colorbox{mygray}{4}[0.15], \colorbox{mygray}{3}[0.07], \colorbox{mygray}{0}[0.06], \colorbox{mygray}{s}[0.05]
}
\\& 
pd.HDF<mask>(
&&&
\\\cdashline{2-5}
&
import numpy as np
&
\multirow{2}{*}{
ar
}
&
\multirow{2}{*}{
\colorbox{pink}{ar}[0.39], \colorbox{mygray}{g}[0.05], \colorbox{mygray}{n}[0.04], \colorbox{mygray}{cal}[0.04], \colorbox{mygray}{b}[0.03]
}
&
\multirow{2}{*}{
\colorbox{mygray}{ ar}[0.94], \colorbox{mygray}{eu}[0.03], \colorbox{pink}{ar}[0.02], \colorbox{mygray}{g}[0.00], \colorbox{mygray}{e}[0.00]
}
\\& 
np.<mask>ccos(
&&&
\\\hline
\multirow{10}{*}{\rotatebox[origin=c]{90}{API Adv. Alias}}

&
import memote.suite.cli.callbacks as attsel
&
\multirow{2}{*}{
git
}
&
\multirow{2}{*}{
\colorbox{mygray}{check}[0.18], \colorbox{mygray}{load}[0.12], \colorbox{mygray}{get}[0.10], \colorbox{mygray}{is}[0.04], \colorbox{mygray}{on}[0.03]
}
&
\multirow{2}{*}{
\colorbox{mygray}{on}[0.27], \colorbox{mygray}{set}[0.07], \colorbox{mygray}{callback}[0.06], \colorbox{mygray}{app}[0.03], \colorbox{mygray}{not}[0.03]
}
\\& 
attsel.<mask>\_installed(
&&&
\\\cdashline{2-5}
&
import weka.flow.conversion as ip
&
\multirow{2}{*}{
Pass
}
&
\multirow{2}{*}{
\colorbox{mygray}{pass}[0.51], \colorbox{mygray}{run}[0.08], \colorbox{pink}{Pass}[0.06], \colorbox{mygray}{connect}[0.02], \colorbox{mygray}{import}[0.02]
}
&
\multirow{2}{*}{
\colorbox{mygray}{pass}[0.69], \colorbox{pink}{Pass}[0.20], \colorbox{mygray}{cycle}[0.06], \colorbox{mygray}{write}[0.01], \colorbox{mygray}{run}[0.01]
}
\\& 
ip.<mask>Through(
&&&
\\\cdashline{2-5}
&
import memote.support.helpers as hermite
&
\multirow{2}{*}{
find
}
&
\multirow{2}{*}{
\colorbox{mygray}{test}[0.27], \colorbox{mygray}{get}[0.13], \colorbox{mygray}{check}[0.09], \colorbox{mygray}{load}[0.05], \colorbox{mygray}{create}[0.04]
}
&
\multirow{2}{*}{
\colorbox{mygray}{start}[0.41], \colorbox{mygray}{stop}[0.15], \colorbox{mygray}{try}[0.04], \colorbox{mygray}{begin}[0.03], \colorbox{mygray}{allow}[0.03]
}
\\& 
hermite.<mask>\_converting\_reactions(
&&&
\\\cdashline{2-5}
&
import pandas as IT
&
\multirow{2}{*}{
Store
}
&
\multirow{2}{*}{
\colorbox{mygray}{5}[0.61], \colorbox{mygray}{S}[0.13], \colorbox{pink}{Store}[0.11], \colorbox{mygray}{4}[0.03], \colorbox{mygray}{Frame}[0.01]
}
&
\multirow{2}{*}{
\colorbox{mygray}{5}[0.27], \colorbox{mygray}{3}[0.12], \colorbox{mygray}{s}[0.11], \colorbox{mygray}{0}[0.10], \colorbox{mygray}{4}[0.07]
}
\\& 
IT.HDF<mask>(
&&&
\\\cdashline{2-5}
&
import numpy as simplejson
&
\multirow{2}{*}{
ar
}
&
\multirow{2}{*}{
\colorbox{mygray}{py}[0.09], \colorbox{mygray}{simple}[0.04], \colorbox{mygray}{g}[0.04], \colorbox{mygray}{cal}[0.03], \colorbox{mygray}{fun}[0.03]
}
&
\multirow{2}{*}{
\colorbox{pink}{ar}[0.28], \colorbox{mygray}{eu}[0.21], \colorbox{mygray}{g}[0.09], \colorbox{mygray}{e}[0.08], \colorbox{mygray}{ce}[0.06]
}
\\& 
simplejson.<mask>ccos(
&&&
\\\hline
    \end{tabular}}
    \caption{Examples of prediction for CodeBERT-MLM-Python and GraphCodeBERT-MLM on API import aliases and their adversarial samples. The last two columns reports the top five tokens generated together with the associated probabilities (in square brackets).}
    \label{tab:rq_alias_example}
\end{table*}


We report $P@K$ results of ten models in Table~\ref{tab:rq_alias} and illustrate examples in Table~\ref{tab:rq_alias_example}. Based on the comparison, CodeBERT-MLM-Python consistently outperforms GraphCodeBERT-MLM, achieving higher $P@1$ scores of up to 24.17\% for Alias scenarios compared to 18.91\%. CodeBERT variants also show better overall performance with $P@50$ scores ranging from 59.88\% to 68.37\%, while GraphCodeBERT-MLM ranges from 59.09\% to 62.17\%. Our initial finding suggests that code models have a weaker understanding of API aliases compared to API calls and imports, as shown in Table~\ref{tab:rq2_table}. This indicates that current code models encode little knowledge of import aliases. Based on the performance of the GPT-3.5-turbo model on 1,000 randomly sampled quizzes, we can infer that it has a greater capability to understand API import aliases. When comparing the results of the original API import alias quizzes with those of the adversarial aliases, we found only minor discrepancies, indicating that these code models have strong robustness in understanding API import aliases. We further analyze the distribution of the API import aliases and find that an API is paired with 1.16 aliases on average, and 8\% of APIs have more than 1 alias. We hypothesize that these code models are able to learn the compositional patterns of these APIs via different aliases, and thus manage to generalize to adversarial import aliases.
\begin{figure*}[!h]
    \centering
    \includegraphics[width=\textwidth]{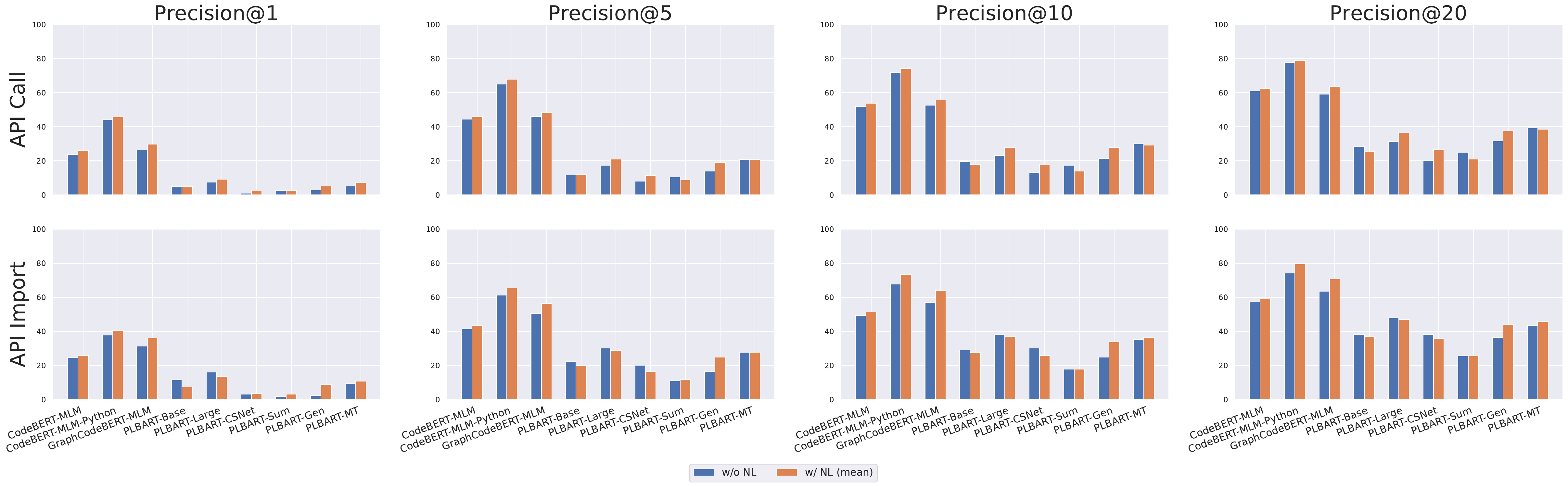}
    \caption{Comparisons of $P@1$, $P@5$, $P@10$ and $P@20$ scores of selected code models. \textbf{w/o NL}: No natural language context is provided along with pop quizzes. \textbf{w/ NL}: Mean performances when natural language context is provided along with pop quizzes.}
    \label{fig:rq3_plot}
\end{figure*}
\finding{of RQ 2}{Although code models show robustness in understanding API import aliases, their encoding of API knowledge is limited to partial information.}

\subsection{RQ3: Will natural language context affect the API name knowledge probing results?}

Our study examines the impact of natural language (NL) context on a code model's ability to comprehend API names. To build a dataset of API-related NL context, we utilize the NL queries designed for Python API sequence usages in the work of~\cite{martin2022deep}. For example, \texttt{<os.path.isfile>} is paired with ``file directory check''. We selected queries containing API sequences containing the \texttt{PyINK} API and transformed them into API pop quizzes. As some of the NL queries are long and infeasible for code models to process, we only choose the 10 shortest ones among these satisfied NL queries, and filter out the cases where the length is no more than 512 tokens determined by the CodeBERT tokenizer. At the end, we collected 56,645 pop quizzes for API calls, and 56,644 for API imports. We compute the average $P@k$ on each API name pop quiz with the group of NL queries concatenated ahead. We compare them with the overall results without adding NL context at the beginning in Figure~\ref{fig:rq3_plot}.

Our results show that incorporating natural language queries led to a 2\% improvement in probing API call name knowledge, although we contend that this may be due to their non-API-focused design, as they were initially created for API sequences rather than individual API calls~\cite{martin2022deep}. In addition, the relative performances among code models do not change before and after adding NL context, suggesting \texttt{PyINK} is robust for API name evaluation. We anticipate that incorporating API-focused natural language context will yield more substantial gains.
\finding{of RQ3}{While adding relevant NL context can aid code models in locating more precise API name knowledge, it does not affect the relative performances among code models in terms of API name knowledge.}

\subsection{RQ4: How well can code models memorize and generalize on API names?}

\begin{table}[!h]
    \centering
    \resizebox{\columnwidth}{!}{
    \begin{tabular}{|l|cc|c|c|c|c|c|c|c|}\hline
    \multicolumn{3}{|c|}{} 
    & \multicolumn{1}{c|}{$P@1 \uparrow$} 
    & \multicolumn{1}{c|}{$P@5 \uparrow$}
    & \multicolumn{1}{c|}{$P@10 \uparrow$} 
    & \multicolumn{1}{c|}{$P@20 \uparrow$}
    & \multicolumn{1}{c|}{$P@30 \uparrow$} 
    & \multicolumn{1}{c|}{$P@40 \uparrow$}
    & \multicolumn{1}{c|}{$P@50 \uparrow$}\\\cline{1-10}

    \multirow{6}{*}{\rotatebox[origin=c]{90}{API Call}}
    &\multirow{2}{*}{CodeBERT}&Seen
& 23.96
& 42.42
& 50.38
& 58.26
& 62.84
& 65.96
& 68.39
\\
&&Unseen
& 21.54
& \textcolor{gray}{\uwave{39.42}}
& \textcolor{gray}{\uwave{46.62}}
& \textcolor{gray}{\uwave{54.92}}
& \textcolor{gray}{\uwave{59.60}}
& \textcolor{gray}{\uwave{62.63}}
& \textcolor{gray}{\uwave{64.35}}
    \\\cdashline{2-10}
    &\multirow{2}{*}{GraphCodeBERT}&Seen
& \textbf{26.00}
& \textbf{43.45}
& \textbf{50.91}
& \textbf{58.45}
& \textbf{62.89}
& \textbf{66.02}
& \textbf{68.42}
\\
&&Unseen
& \textcolor{gray}{\uwave{21.86}}
& 37.42
& 44.44
& 51.89
& 56.97
& 60.26
& 62.36

    \\\cdashline{2-10}
    &\multirow{2}{*}{PLBART}&Seen
& 2.83
& 11.24
& 16.09
& 22.08
& 26.20
& 29.45
& 32.10
\\
&&Unseen
& 2.00
& 8.31
& 12.80
& 18.38
& 21.94
& 24.52
& 27.32
    \\\hline\hline
    
    \multirow{6}{*}{\rotatebox[origin=c]{90}{API Import}}
    &\multirow{2}{*}{CodeBERT}&Seen
& 26.49
& 44.01
& 51.34
& 58.80
& 63.06
& 66.03
& 68.35
    \\
    &&Unseen
& 20.46
& \textcolor{gray}{\uwave{38.83}}
& \textcolor{gray}{\uwave{45.82}}
& 52.58
& 56.49
& 58.94
& 61.08
    \\\cdashline{2-10}
    &\multirow{2}{*}{GraphCodeBERT}&Seen
& \textbf{30.99}
& \textbf{48.51}
& \textbf{55.61}
& \textbf{62.74}
& \textbf{66.92}
& \textbf{69.79}
& \textbf{71.96}
    \\
    &&Unseen
& \textcolor{gray}{\uwave{22.37}}
& 38.41
& 45.81
& \textcolor{gray}{\uwave{53.69}}
& \textcolor{gray}{\uwave{57.83}}
& \textcolor{gray}{\uwave{60.78}}
& \textcolor{gray}{\uwave{63.16}}

    \\\cdashline{2-10}
    &\multirow{2}{*}{PLBART}&Seen
& 3.64
& 15.29
& 22.39
& 30.80
& 36.29
& 40.37
& 43.63
    \\
    &&Unseen
& 2.17
& 9.96
& 16.42
& 23.97
& 29.08
& 32.59
& 35.65

    \\\hline
    \end{tabular}}
    \caption{Comparison of $P@k$ scores on \textbf{seen} and \textbf{unseen} API name quizzes among selected code models.}
    \label{tab:rq4}
\end{table}
We evaluate whether code models demonstrate a deeper understanding of the names of \textbf{seen} APIs during pre-training than the \textbf{unseen} ones by conducting experiments on CodeBERT-MLM, GraphCodeBERT-MLM, and PLBART-CSNet, which were pre-trained on the train set of CSNet. To create our 
 \texttt{PyINK-Mem} version, we take all APIs appearing during training as the \textbf{seen} split and the remaining APIs as the \textbf{unseen} split. We filter out all the APIs belonging to the \textbf{seen} libraries in the \textbf{unseen} split. The \texttt{PyINK-Mem} \textbf{seen} split contained 281,945 quizzes for API calls and 281,945 quizzes for API imports. The unseen split had 7,640 API call quizzes and 7,640 API import quizzes. We note that the selected models do not memorize any structures of the open-source repositories, due to the function-level pre-training objective.

 In Table~\ref{tab:rq4}, we measure the model performance via $P@k$ up to $P@50$. Our inspection of the results on API call pop quizzes suggests there are slight differences between \textbf{seen} and \textbf{unseen} sets, indicating the strong generalization ability of these code models to new APIs.  Among the three models we evaluated, CodeBERT-MLM demonstrates the most robust performance, while GraphCodeBERT-MLM demonstrates a greater ability to memorize API names during pre-training. Surprisingly, we found that there were 1,288 and 5,468 distinct ground-truth tokens in the \textbf{seen} and \textbf{unseen} splits for API calls, respectively, and 1,257 tokens (97.59\% of the \textbf{unseen} split) were overlapped. This indicates that the API namespace designs share unexpected commonalities.

 \finding{of RQ4}{Code models demonstrate impressive generalization abilities in predicting the names of programming functions for new domains and reasonable memorizations of APIs from the training data.}

\section{Discussion}

This section details the implications and importance of our experiments, and discusses threats to their validity.

\subsection{The implication of the evaluation on \texttt{PyINK}}
Our study evaluates API name knowledge in code models on the \texttt{PyINK} benchmark, encompassing a wide range of Python APIs. Our analysis reveals that while these APIs share similar design patterns regarding fully qualified naming, they also exhibit diverse characteristics. This observation underscores the challenges associated with modeling and comprehending the name information pertaining to these APIs. Our experiments on \texttt{PyINK} provide qualitative insight into the API name knowledge in the pre-trained code models. However, it is difficult to infer the specific quantitative amount, given the models are selected based on different architectures. To address this, we suggest comparing models from the same family would be more reasonable.

Based on a detailed evaluation of \texttt{PyINK}, we find that code models can store some API name knowledge while learning from source code, but the knowledge is often insufficient. Although these models achieve reasonable precision within the top-50 predictions, they may struggle to answer pop quizzes correctly within the first few attempts. One consistent finding across various experiments is that code models may not treat API calls and imports similarly, even though these two are derived from the same API fully qualified name.

We also empirically demonstrate that the code model's ability to capture API name knowledge can be influenced by several factors, including pre-training strategies, model parameters, pre-training data, and fine-tuning. Our results from RQ3 indicate that natural language queries can assist code models in locating specific API name information to some extent, underscoring the importance of context in constructing the approach. Furthermore, we emphasize that API name knowledge is distinct from most knowledge in natural language, where code models can generalize to unseen API names. In comparison, language models consistently perform inferior when generalizing to new domains~\cite{tanzer2021memorisation,carlini2022quantifying}.

\subsection{Strengths of \texttt{INK} evaluation framework}

The \texttt{INK} framework has two key strengths. Firstly, it can diagnose code models' understanding of API names. While we found that discrete prompts may not be sufficient for instructing models like GPT-3.5 to complete pop quizzes, we recommend using few-shot examples in prompts to improve performance. For publicly generative code models such as CodeGen~\cite{nijkamp2023codegen} and SantaCoder~\cite{allal2023santacoder} that may not be able to perform API name understanding tasks directly, we can use prompt tuning strategies like P-tuning~\cite{liu2022p} to effectively evaluate their performance.

Secondly, \texttt{INK}'s cloze-style pop quiz on API name understanding can be a valuable pre-training and fine-tuning task for code models. By learning to predict API names, code models can develop better representations of APIs and can be more easily adapted to various downstream tasks involving APIs, such as code summarization on API invocation and invoking APIs during code completion.

\subsection{The importance of interpreting API name knowledge}

Acquiring knowledge of API names is essential for software development, as APIs play a critical role in automating various functionalities. However, code models used in current benchmarks do not adequately consider API usage when assessing performance. Our research shows that code models often struggle to preserve essential API name knowledge, leading to the need for knowledge-enhanced training paradigms. Inspired by recent advancements in knowledge-enhanced pre-trained models in natural language processing and computer vision, we propose exploring approaches that integrate API knowledge graphs into code models. Learning from research on API knowledge graphs~\cite{li2018improving,ren2020api,wang2021novel} holds great promise for advancing the field. By incorporating these approaches into code models, we can improve the accuracy and effectiveness of automated software development.

\subsection{Threats to Validity}

\paragraph{Choice of Cloze-style Pop Quiz Design}
The employed pop quiz design involves predicting the partial or full name of each API module level. It is possible to prompt code models to predict on the tokens in between when an API level involves multiple tokens. However, masking the first and last tokens of each module level is a suitable approach for assessing the extent of API name knowledge retained by code models, validating our pop quiz choice. While some may criticize our pop quizzes for targeting single token prediction in the missing slot instead of multiple tokens, this design choice aligns with prior works in probing language models~\cite{petroni2019language}. Introducing multi-token prediction could lead to more variability in decoding strategies among code models and make it difficult to finalize the sequence combined by all predicted tokens during evaluation. Although we acknowledge this potential threat to validity, we leave it to future work.
\paragraph{Choice of Evaluation Data}

The choice of evaluation data may affect experimental results of \texttt{INK} distillation. While we used a widely-used corpus, CSNet, which covers a substantial number of Python APIs, it is important to acknowledge that there are additional resources, such as Stack Overflow\footnote{\url{https://stackoverflow.com/}}, that may contain more Python APIs. Moreover, CSNet was proposed in 2019, and new APIs may have been developed since then. We contend that our evaluation of \texttt{PyINK} using CSNet is statistically significant, but we also acknowledge the limitations of this corpus. Furthermore, code models may exhibit different behaviors when evaluated with APIs in other programming languages such as Java and C. To address this threat to validity, we can enhance the completeness of our evaluation by incorporating more programming languages on which these code models are trained. By evaluating the code models on a wider range of programming languages, we can better ensure their robustness and generalizability to real-world programming tasks.



\section{Conclusion}
In this paper, we have explored the interpretability of code models for source code (CodeBERT, GraphCodeBERT and PLBART). We conduct a thorough API name knowledge analysis based on a large-scale benchmark, \texttt{PyINK}, from the following four aspects, aiming to give an interpretation of 
code models. Firstly, we determine the API name knowledge stored by code models from two perspectives, API call and API import. Secondly, we investigate whether code models can robustly understand API import aliases. Thirdly, we revisit the settings in deep API learning and assess if providing additional natural language context can help code model retrieve more precise API name knowledge. Fourthly, we examine the memorization and generalization of code models on API names. The analysis in this paper has revealed several interesting findings that can inspire future studies on code representation learning and interpretation of knowledge encoded by code models.

\bibliographystyle{IEEEtran}
\bibliography{ref}

\begin{thebibliography}{10}
\providecommand{\url}[1]{#1}
\csname url@samestyle\endcsname
\providecommand{\newblock}{\relax}
\providecommand{\bibinfo}[2]{#2}
\providecommand{\BIBentrySTDinterwordspacing}{\spaceskip=0pt\relax}
\providecommand{\BIBentryALTinterwordstretchfactor}{4}
\providecommand{\BIBentryALTinterwordspacing}{\spaceskip=\fontdimen2\font plus
\BIBentryALTinterwordstretchfactor\fontdimen3\font minus
  \fontdimen4\font\relax}
\providecommand{\BIBforeignlanguage}[2]{{%
\expandafter\ifx\csname l@#1\endcsname\relax
\typeout{** WARNING: IEEEtran.bst: No hyphenation pattern has been}%
\typeout{** loaded for the language `#1'. Using the pattern for}%
\typeout{** the default language instead.}%
\else
\language=\csname l@#1\endcsname
\fi
#2}}
\providecommand{\BIBdecl}{\relax}
\BIBdecl

\bibitem{feng2020codebert}
Z.~Feng, D.~Guo, D.~Tang, N.~Duan, X.~Feng, M.~Gong, L.~Shou, B.~Qin, T.~Liu,
  D.~Jiang \emph{et~al.}, ``Codebert: A pre-trained model for programming and
  natural languages,'' in \emph{Findings of the Association for Computational
  Linguistics: EMNLP 2020}, 2020, pp. 1536--1547.

\bibitem{guo2021graphcodebert}
\BIBentryALTinterwordspacing
D.~Guo, S.~Ren, S.~Lu, Z.~Feng, D.~Tang, S.~LIU, L.~Zhou, N.~Duan,
  A.~Svyatkovskiy, S.~Fu, M.~Tufano, S.~K. Deng, C.~Clement, D.~Drain,
  N.~Sundaresan, J.~Yin, D.~Jiang, and M.~Zhou, ``Graphcode{\{}bert{\}}:
  Pre-training code representations with data flow,'' in \emph{International
  Conference on Learning Representations}, 2021. [Online]. Available:
  \url{https://openreview.net/forum?id=jLoC4ez43PZ}
\BIBentrySTDinterwordspacing

\bibitem{allamanis2018survey}
M.~Allamanis, E.~T. Barr, P.~Devanbu, and C.~Sutton, ``A survey of machine
  learning for big code and naturalness,'' \emph{ACM Computing Surveys (CSUR)},
  vol.~51, no.~4, pp. 1--37, 2018.

\bibitem{10.1145/3487569}
\BIBentryALTinterwordspacing
F.~F. Xu, B.~Vasilescu, and G.~Neubig, ``In-ide code generation from natural
  language: Promise and challenges,'' \emph{ACM Trans. Softw. Eng. Methodol.},
  vol.~31, no.~2, mar 2022. [Online]. Available:
  \url{https://doi.org/10.1145/3487569}
\BIBentrySTDinterwordspacing

\bibitem{niudeep}
C.~Niu, C.~Li, B.~Luo, and V.~Ng, ``Deep learning meets software engineering: A
  survey on pre-trained models of source code.''

\bibitem{chen2021evaluating}
M.~Chen, J.~Tworek, H.~Jun, Q.~Yuan, H.~P. d.~O. Pinto, J.~Kaplan, H.~Edwards,
  Y.~Burda, N.~Joseph, G.~Brockman \emph{et~al.}, ``Evaluating large language
  models trained on code,'' \emph{arXiv preprint arXiv:2107.03374}, 2021.

\bibitem{li2023starcoder}
R.~Li, L.~B. Allal, Y.~Zi, N.~Muennighoff, D.~Kocetkov, C.~Mou, M.~Marone,
  C.~Akiki, J.~Li, J.~Chim \emph{et~al.}, ``Starcoder: may the source be with
  you!'' \emph{arXiv preprint arXiv:2305.06161}, 2023.

\bibitem{openai2023gpt4}
OpenAI, ``Gpt-4 technical report,'' 2023.

\bibitem{lai2022ds}
Y.~Lai, C.~Li, Y.~Wang, T.~Zhang, R.~Zhong, L.~Zettlemoyer, S.~W.-t. Yih,
  D.~Fried, S.~Wang, and T.~Yu, ``Ds-1000: A natural and reliable benchmark for
  data science code generation,'' \emph{arXiv preprint arXiv:2211.11501}, 2022.

\bibitem{yu2023codereval}
H.~Yu, B.~Shen, D.~Ran, J.~Zhang, Q.~Zhang, Y.~Ma, G.~Liang, Y.~Li, T.~Xie, and
  Q.~Wang, ``Codereval: A benchmark of pragmatic code generation with
  generative pre-trained models,'' \emph{arXiv preprint arXiv:2302.00288},
  2023.

\bibitem{samek2017explainable}
W.~Samek, T.~Wiegand, and K.-R. M{\"u}ller, ``Explainable artificial
  intelligence: Understanding, visualizing and interpreting deep learning
  models,'' \emph{arXiv preprint arXiv:1708.08296}, 2017.

\bibitem{jiang2020can}
Z.~Jiang, F.~F. Xu, J.~Araki, and G.~Neubig, ``How can we know what language
  models know?'' \emph{Transactions of the Association for Computational
  Linguistics}, vol.~8, pp. 423--438, 2020.

\bibitem{wan2022they}
Y.~Wan, W.~Zhao, H.~Zhang, Y.~Sui, G.~Xu, and H.~Jin, ``What do they capture? a
  structural analysis of pre-trained language models for source code,'' in
  \emph{Proceedings of the 44th International Conference on Software
  Engineering}, 2022, pp. 2377--2388.

\bibitem{hernandez2022ast}
J.~A. Hern{\'a}ndez~L{\'o}pez, M.~Weyssow, J.~S. Cuadrado, and H.~Sahraoui,
  ``Ast-probe: Recovering abstract syntax trees from hidden representations of
  pre-trained language models,'' in \emph{37th IEEE/ACM International
  Conference on Automated Software Engineering}, 2022, pp. 1--11.

\bibitem{troshin2022probing}
S.~Troshin and N.~Chirkova, ``Probing pretrained models of source code,''
  \emph{arXiv preprint arXiv:2202.08975}, 2022.

\bibitem{huang2022prompt}
Q.~Huang, Z.~Yuan, Z.~Xing, X.~Xu, L.~Zhu, and Q.~Lu, ``Prompt-tuned code
  language model as a neural knowledge base for type inference in
  statically-typed partial code,'' in \emph{37th IEEE/ACM International
  Conference on Automated Software Engineering}, 2022, pp. 1--13.

\bibitem{ahmad2021unified}
W.~Ahmad, S.~Chakraborty, B.~Ray, and K.-W. Chang, ``Unified pre-training for
  program understanding and generation,'' in \emph{Proceedings of the 2021
  Conference of the North American Chapter of the Association for Computational
  Linguistics: Human Language Technologies}, 2021, pp. 2655--2668.

\bibitem{devlin2019bert}
J.~Devlin, M.-W. Chang, K.~Lee, and K.~Toutanova, ``Bert: Pre-training of deep
  bidirectional transformers for language understanding,'' in \emph{Proceedings
  of the 2019 Conference of the North American Chapter of the Association for
  Computational Linguistics: Human Language Technologies, Volume 1 (Long and
  Short Papers)}, 2019, pp. 4171--4186.

\bibitem{qiu2020pre}
X.~Qiu, T.~Sun, Y.~Xu, Y.~Shao, N.~Dai, and X.~Huang, ``Pre-trained models for
  natural language processing: A survey,'' \emph{Science China Technological
  Sciences}, vol.~63, no.~10, pp. 1872--1897, 2020.

\bibitem{hindle2016naturalness}
A.~Hindle, E.~T. Barr, M.~Gabel, Z.~Su, and P.~Devanbu, ``On the naturalness of
  software,'' \emph{Communications of the ACM}, vol.~59, no.~5, pp. 122--131,
  2016.

\bibitem{husain2019codesearchnet}
H.~Husain, H.-H. Wu, T.~Gazit, M.~Allamanis, and M.~Brockschmidt,
  ``Codesearchnet challenge: Evaluating the state of semantic code search,''
  \emph{arXiv preprint arXiv:1909.09436}, 2019.

\bibitem{lewis2020bart}
M.~Lewis, Y.~Liu, N.~Goyal, M.~Ghazvininejad, A.~Mohamed, O.~Levy, V.~Stoyanov,
  and L.~Zettlemoyer, ``Bart: Denoising sequence-to-sequence pre-training for
  natural language generation, translation, and comprehension,'' in
  \emph{Proceedings of the 58th Annual Meeting of the Association for
  Computational Linguistics}, 2020, pp. 7871--7880.

\bibitem{liu2023pre}
P.~Liu, W.~Yuan, J.~Fu, Z.~Jiang, H.~Hayashi, and G.~Neubig, ``Pre-train,
  prompt, and predict: A systematic survey of prompting methods in natural
  language processing,'' \emph{ACM Computing Surveys}, vol.~55, no.~9, pp.
  1--35, 2023.

\bibitem{petroni2019language}
F.~Petroni, T.~Rockt{\"a}schel, S.~Riedel, P.~Lewis, A.~Bakhtin, Y.~Wu, and
  A.~Miller, ``Language models as knowledge bases?'' in \emph{Proceedings of
  the 2019 Conference on Empirical Methods in Natural Language Processing and
  the 9th International Joint Conference on Natural Language Processing
  (EMNLP-IJCNLP)}, 2019, pp. 2463--2473.

\bibitem{jiang2020x}
Z.~Jiang, A.~Anastasopoulos, J.~Araki, H.~Ding, and G.~Neubig, ``X-factr:
  Multilingual factual knowledge retrieval from pretrained language models,''
  in \emph{Proceedings of the 2020 Conference on Empirical Methods in Natural
  Language Processing (EMNLP)}, 2020, pp. 5943--5959.

\bibitem{petroni2020context}
\BIBentryALTinterwordspacing
F.~Petroni, P.~Lewis, A.~Piktus, T.~Rockt{\"a}schel, Y.~Wu, A.~H. Miller, and
  S.~Riedel, ``How context affects language models' factual predictions,'' in
  \emph{Automated Knowledge Base Construction}, 2020. [Online]. Available:
  \url{https://openreview.net/forum?id=025X0zPfn}
\BIBentrySTDinterwordspacing

\bibitem{perez2021true}
E.~Perez, D.~Kiela, and K.~Cho, ``True few-shot learning with language
  models,'' \emph{Advances in neural information processing systems}, vol.~34,
  pp. 11\,054--11\,070, 2021.

\bibitem{qin2021learning}
G.~Qin and J.~Eisner, ``Learning how to ask: Querying lms with mixtures of soft
  prompts,'' in \emph{Proceedings of the 2021 Conference of the North American
  Chapter of the Association for Computational Linguistics: Human Language
  Technologies}, 2021, pp. 5203--5212.

\bibitem{huang2018api}
Q.~Huang, X.~Xia, Z.~Xing, D.~Lo, and X.~Wang, ``Api method recommendation
  without worrying about the task-api knowledge gap,'' in \emph{Proceedings of
  the 33rd ACM/IEEE International Conference on Automated Software
  Engineering}, 2018, pp. 293--304.

\bibitem{rahman2016rack}
M.~M. Rahman, C.~K. Roy, and D.~Lo, ``Rack: Automatic api recommendation using
  crowdsourced knowledge,'' in \emph{2016 IEEE 23rd International Conference on
  Software Analysis, Evolution, and Reengineering (SANER)}, vol.~1.\hskip 1em
  plus 0.5em minus 0.4em\relax IEEE, 2016, pp. 349--359.

\bibitem{thung2013automatic}
F.~Thung, S.~Wang, D.~Lo, and J.~Lawall, ``Automatic recommendation of api
  methods from feature requests,'' in \emph{2013 28th IEEE/ACM International
  Conference on Automated Software Engineering (ASE)}.\hskip 1em plus 0.5em
  minus 0.4em\relax IEEE, 2013, pp. 290--300.

\bibitem{gu2016deep}
X.~Gu, H.~Zhang, D.~Zhang, and S.~Kim, ``Deep api learning,'' in
  \emph{Proceedings of the 2016 24th ACM SIGSOFT international symposium on
  foundations of software engineering}, 2016, pp. 631--642.

\bibitem{martin2022deep}
J.~Martin and J.~L. Guo, ``Deep api learning revisited,'' in \emph{Proceedings
  of the 30th IEEE/ACM International Conference on Program Comprehension},
  2022, pp. 321--330.

\bibitem{hadi2022effectiveness}
M.~A. Hadi, I.~N.~B. Yusuf, F.~Thung, K.~G. Luong, J.~Lingxiao, F.~H. Fard, and
  D.~Lo, ``On the effectiveness of pretrained models for api learning,'' in
  \emph{Proceedings of the 30th IEEE/ACM International Conference on Program
  Comprehension}, 2022, pp. 309--320.

\bibitem{papineni2002bleu}
K.~Papineni, S.~Roukos, T.~Ward, and W.-J. Zhu, ``Bleu: a method for automatic
  evaluation of machine translation,'' in \emph{Proceedings of the 40th annual
  meeting of the Association for Computational Linguistics}, 2002, pp.
  311--318.

\bibitem{wang2021restoring}
J.~Wang, L.~Li, and A.~Zeller, ``Restoring execution environments of jupyter
  notebooks,'' in \emph{2021 IEEE/ACM 43rd International Conference on Software
  Engineering (ICSE)}.\hskip 1em plus 0.5em minus 0.4em\relax IEEE, 2021, pp.
  1622--1633.

\bibitem{wolf2020transformers}
T.~Wolf, L.~Debut, V.~Sanh, J.~Chaumond, C.~Delangue, A.~Moi, P.~Cistac,
  T.~Rault, R.~Louf, M.~Funtowicz \emph{et~al.}, ``Transformers:
  State-of-the-art natural language processing,'' in \emph{Proceedings of the
  2020 conference on empirical methods in natural language processing: system
  demonstrations}, 2020, pp. 38--45.

\bibitem{aghajanyan2023scaling}
A.~Aghajanyan, L.~Yu, A.~Conneau, W.-N. Hsu, K.~Hambardzumyan, S.~Zhang,
  S.~Roller, N.~Goyal, O.~Levy, and L.~Zettlemoyer, ``Scaling laws for
  generative mixed-modal language models,'' \emph{arXiv preprint
  arXiv:2301.03728}, 2023.

\bibitem{tanzer2021memorisation}
M.~T{\"a}nzer, S.~Ruder, and M.~Rei, ``Memorisation versus generalisation in
  pre-trained language models,'' \emph{arXiv preprint arXiv:2105.00828}, 2021.

\bibitem{carlini2022quantifying}
\BIBentryALTinterwordspacing
N.~Carlini, D.~Ippolito, M.~Jagielski, K.~Lee, F.~Tramer, and C.~Zhang,
  ``Quantifying memorization across neural language models,'' in \emph{The
  Eleventh International Conference on Learning Representations}, 2023.
  [Online]. Available: \url{https://openreview.net/forum?id=TatRHT_1cK}
\BIBentrySTDinterwordspacing

\bibitem{nijkamp2023codegen}
\BIBentryALTinterwordspacing
E.~Nijkamp, B.~Pang, H.~Hayashi, L.~Tu, H.~Wang, Y.~Zhou, S.~Savarese, and
  C.~Xiong, ``Codegen: An open large language model for code with multi-turn
  program synthesis,'' in \emph{The Eleventh International Conference on
  Learning Representations}, 2023. [Online]. Available:
  \url{https://openreview.net/forum?id=iaYcJKpY2B_}
\BIBentrySTDinterwordspacing

\bibitem{allal2023santacoder}
L.~B. Allal, R.~Li, D.~Kocetkov, C.~Mou, C.~Akiki, C.~M. Ferrandis,
  N.~Muennighoff, M.~Mishra, A.~Gu, M.~Dey \emph{et~al.}, ``Santacoder: don't
  reach for the stars!'' \emph{arXiv preprint arXiv:2301.03988}, 2023.

\bibitem{liu2022p}
X.~Liu, K.~Ji, Y.~Fu, W.~Tam, Z.~Du, Z.~Yang, and J.~Tang, ``P-tuning: Prompt
  tuning can be comparable to fine-tuning across scales and tasks,'' in
  \emph{Proceedings of the 60th Annual Meeting of the Association for
  Computational Linguistics (Volume 2: Short Papers)}, 2022, pp. 61--68.

\bibitem{li2018improving}
H.~Li, S.~Li, J.~Sun, Z.~Xing, X.~Peng, M.~Liu, and X.~Zhao, ``Improving api
  caveats accessibility by mining api caveats knowledge graph,'' in \emph{2018
  IEEE International Conference on Software Maintenance and Evolution
  (ICSME)}.\hskip 1em plus 0.5em minus 0.4em\relax IEEE, 2018, pp. 183--193.

\bibitem{ren2020api}
X.~Ren, X.~Ye, Z.~Xing, X.~Xia, X.~Xu, L.~Zhu, and J.~Sun, ``Api-misuse
  detection driven by fine-grained api-constraint knowledge graph,'' in
  \emph{Proceedings of the 35th IEEE/ACM International Conference on Automated
  Software Engineering}, 2020, pp. 461--472.

\bibitem{wang2021novel}
X.~Wang, X.~Liu, J.~Liu, X.~Chen, and H.~Wu, ``A novel knowledge graph
  embedding based api recommendation method for mashup development,''
  \emph{World Wide Web}, vol.~24, no.~3, pp. 869--894, 2021.

\end{thebibliography}

\end{document}